# Location and Spectral Estimation of Weak Wave Packets on Noise Background

## Yu. Bunyak[1] and O. Bunyak[2]

[1]InnoVinn Inc., Kievskaya 14, Vinnitsa, Ukraine, 21001, E-mail: yuri.bunyak@innovinn.com
[2]Vinnitsa National Technical University, Chmelnitsky rode 75, 21021, Vinnitsa, Ukraine

*Abstract.* The method of location and spectral estimation of weak signals on a noise background is being considered. The method is based on the optimized on order and noise dispersion autoregressive model of a sought signal. A new approach of model order determination is being offered. Available estimation of the noise dispersion is close to the real one. The optimized model allows to define function of empirical data spectral and dynamic features changes. The analysis of the signal as dynamic invariant in respect of the linear shift transformation yields the function of model consistency. Use of these both functions enables to detect short-time and nonstationary wave packets at signal to noise ratio as from -20 dB and above.

## 1. Introduction

A problem of finding weak signals with unknown parameters on a noise background is important for many physical and technical investigations. Usually the signals are different types of fluctuations with expressed deterministic dynamic characteristic. The methods that are used to solve the problems of signals finding are based on analysis of the statistical characteristics of background noise and signal [1,2]. If statistical features of the signal and the noise are characterized as a constant during some time interval then it is possible to use well known optimum and adaptive methods of signal reception [1,2,3]. Some of the methods consider events of the signals form abrupt changes basing on the given statistical observations [4]. In many cases it is impossible to note some changes of the statistics since signal doesn't have well-marked sidebars and shows itself along short time only, or it has varying parameters. The mentioned methods are not capable to find the moment of dynamic change and the signal under such conditions, particularly at low signal to noise ratio (SNR). Described signals are peculiar to the problems of finding gravitational and seismic waves, parameters measurement of electromagnetic fluctuations in plasma and objects of biophysics, vibrations of mechanical systems.

In present article the method of finding short-time harmonic packets and chirp-signals on noise background is considered. The method is based on the moving correlation and autoregressive analysis using a short-time interval. The time interval is chosen with the assumption that signal during the interval is ergodic, stationary and its dynamic can be characterized by autoregressive (AR) model [5]

$$x_n = -\sum_{i=1}^{p} a_i x_{n-i} + \xi_n, \qquad (1)$$

where

$$x_n = s_n + \varsigma_n \qquad (2)$$

are data samples which possible to present as a sum (2) of the harmonic signal $s_n$ and the noise $\varsigma_n$ with zero mean and the dispersion $\sigma_\varsigma^2$, $a_i$ – coefficients (parameters) of the linear prediction (LP), $p$ – order of the autoregression (AR), $\xi_n$ – samples of the white noise with the dispersion $\sigma_\xi^2$, $n=0,1,...$.

We shall call attention that the model AR (1) is used for description both the sum of harmonic signals on noise background and stochastic stationary processes. In the first case, the roots of the characteristic polynomial of the AR are connected with frequencies and damping factors of harmonics [5,6], samples of the noise are a component part of the signal, as it presented in equation (2). The problem of AR parameters estimation must be set from the condition of $\min(\sigma_\xi^2 - \sigma_\varsigma^2)$. In the second case, the roots of the characteristic polynomial are located inside the unit circle of the complex area, samples of the white noise present an external innovation process, which power compensates contributed by LP damping. The problem of AR parameters estimation must be set from condition of $\min(\sigma_\xi^2)$. The first criterion must be chosen for solution the problem of signals finding, since in this case value of the function

$$G = (P_x - \sigma_\xi^2) > G_{\min}, \qquad (3)$$

where $P_x$ – average power of data $x_n$ on the interval of analysis, $G_{\min}$ – some threshold value, can be used as a criterion of the signal $s_n$ presence or absences. However, primary majority of the methods of AR parameters estimation use the second criterion [5,6]. For AR parameters estimation according to the first criterion it is possible to refer the methods that are based on the presentation of data or correlation matrixes in signal or noise subspaces using eigenvectors decomposition or singular values decompositions [6]. We will consider this problem from standpoint of data dynamic characteristic analysis by the method of sought signal $s_n$ invariants.

## 2. AR parameters estimation

We will write the expression for the AR process (1) as a LP matrix equation for signal component:

$$\mathbf{S}_1 = \mathbf{K}_p \mathbf{S}_0, \qquad (4)$$

where $\mathbf{S}_{0(1)} = [s_{i+k(+1)}]_{i=0...p-1, k=0(1)...N-p-1(+1)}$, $N$ - some number of samples during time interval that is comparable to duration of signal, the matrix

$$\mathbf{K}_p = \begin{bmatrix} 0 & 1 & \cdots & 0 \\ \vdots & \vdots & \ddots & \vdots \\ 0 & 0 & \cdots & 1 \\ -a_p & -a_{p-1} & \cdots & -a_1 \end{bmatrix} \qquad (5)$$

is known as Frobenius matrix, its eigenvalues equal to the roots of the characteristic polynomial of the LP model (4). The matrix (5) plays the role of the linear shift operator (LSO) [7] in equation (4). The correlation matrix of the signal is possible to estimate as $\mathbf{R}_s = \mathbf{S}_0 \mathbf{S}_0^H$, where $H$ - hermitian conjugate, or as $\mathbf{R}_s' = \mathbf{S}_1 \mathbf{S}_1^H$. It is possible to expect that $\mathbf{R}_s = \mathbf{R}_s'$ if signal is stationary and taking into consideration (4)

$$\mathbf{R}_s = \mathbf{K}_p \mathbf{R}_s \mathbf{K}_p^H. \tag{6}$$

The equation (6) defines the characteristic of linear symmetry (LS) of the correlation matrix of harmonic signal. Transformation of the correlation matrix in the right part of the expression (6) brings to the shift of matrix elements along its diagonal to the left and complements its last row and column by elements, which are formed by equation of the LP and the last diagonal element by quadratic form with coefficients $a_i$. Therefore the correlation matrix with Toeplitz structure can possess the characteristic of LS (6). Estimation of such matrix on a short interval presents a difficult problem [8], so instead of equation (6) we will use the maximum likelihood (MLH) function of data correlation matrixes $\mathbf{R}_x$ and $\mathbf{R}_x'$ that are similar to $\mathbf{R}_s$ and $\mathbf{R}_s'$ [8]:

$$g(\mathbf{R}_x, \mathbf{R}_x') = -\ln(\det \mathbf{R}_x) - tr(\mathbf{R}_x^{-1} \mathbf{R}_x'). \tag{7}$$

We will find AR parameters from extremum condition of the MLH function (7):

$$\frac{\partial}{\partial a_i} g(\mathbf{R}_x^{-1} \mathbf{R}_x') = 0;\ i = 1, \ldots, p. \tag{8}$$

Since the first component in the right part of equation (7) does not depend on AR parameters the equation (8) is equivalent to following one [9]:

$$\frac{\partial}{\partial a_i} tr(\mathbf{R}_x^{-1} \mathbf{K}_p \mathbf{R}_x \mathbf{K}_p^H) = 0;\ i = 1, \ldots, p. \tag{9}$$

The equation (9) has the solution in the manner of the LP parameters vector

$$a_i = \rho_{p-1-i, p-1} \rho_{p-1, p-1}^{-1};\ i = 1, \ldots, p-1;\ a_p = 0, \tag{10}$$

where $\rho_{\cdot, p-1}$ are elements of the last column of the matrix $\mathbf{R}_x^{-1}$, let us denote it as $\rho_{(p-1)}$. With account of (10) the equation (9) is equivalent to the matrix expression

$$\begin{bmatrix} r_{0,0} & r_{0,1} & \cdots & r_{0,p-1} \\ r_{1,0} & r_{1,1} & \cdots & r_{1,p-1} \\ \vdots & \vdots & \ddots & \vdots \\ r_{p-1,0} & r_{p-1,1} & \cdots & r_{p-1,p-1} \end{bmatrix} \cdot \begin{bmatrix} a_{p-1} \\ a_{p-2} \\ \vdots \\ 1 \end{bmatrix} = \begin{bmatrix} 0 \\ 0 \\ \vdots \\ \rho_{p-1,p-1}^{-1} \end{bmatrix}, \tag{11}$$

where $r_{i,k}$ are elements of the matrix $\mathbf{R}_x$. Comparing the equation (11) with known equations on the basis of the method of the least squares [5,6] we can conclude that estimation of the noise dispersion in the expression (1) can be defined as

$$\sigma_\xi^2 = \rho_{p-1, p-1}^{-1}. \tag{12}$$

The dispersion (12) does not depend on parameters of LP and is defined only by correlation characteristic of data. Unlike equations that are based on the least squares, the dispersion in equation (11) is known and is obtained from condition of signal invariance in respect of dynamical transformation by the shift operator (5). It only depends on the order of the model. Essential influence also can be rendered by the method of correlation matrix estimation. Statistical sense of the estimation (12) consists in well known condition that correlation matrix of a stationary AR process is its Fisher's information matrix (FIM) [10]. Diagonal elements of the inverse to FIM matrix install the limiting accuracy of AR parameters estimation that is known as Kramer-Rao's bound (KRB). Therefore, the dispersion of the estimation error $\sigma_\xi^2$ is equal to dispersion of the noise in (2), consequently $\sigma_\xi^2 \approx \sigma_\varsigma^2$. Under such condition the function (3) can be used as criterion of signal finding on noise background. However, it is also necessary to have a criterion that found signal is a regular fluctuation. This criterion can be received from statistical analysis of signal invariance with respect to shift transformation by LSO.

### 3. Function of the dynamical invariance criterion

We will denote the LP parameters vector as $\mathbf{a} = [a_i]_{i=1 \ldots p}$. From characteristic of signal invariance with respect to the shift transformation (4) follows that the matrixes $\mathbf{S}_m = \mathbf{K}_p^m \mathbf{S}_0$ can be formed, where $m = 1, 2, \ldots$. Using two of them, when $m = 1$ and $m = 2$, we can write the following equations for arbitrary sample of data $x_n$:

$$x_n = -\sum_{i=1}^{p} a_i x_{n-i} + \xi_n^{(1)}, \tag{13.1}$$

$$x_n = -\sum_{i=1}^{p} (a_{i+1} - a_1 a_i) x_{n-i-1} + \xi_n^{(2)}, \tag{13.2}$$

where $a_{p+1} = 0$, vector of parameters in equation (13.2) is determined as $\mathbf{a}^T \mathbf{K}_p$, $T$ - transposition. For analysis of noise influence we use KRB. It is well known that KRB does not depend on the method, which used for estimation of AR parameters, it depends on probability density function (PDF) of the noise. If we consider the equations (13) as a system, then PDF of noise will be changed. We will take simplifying admissions that $\xi_n^{(1)}$ and $\xi_n^{(2)}$ are statistically independent and have joint conditional Gauss PDF [10] $P(\xi_n | \mathbf{a}) = P(\xi_n^{(1)} | \mathbf{a}) P(\xi_n^{(2)} | \mathbf{a}^T \mathbf{K}_p)$. Let us evaluate FIM as

$$\mathbf{F} = -\left[ \prod_{n=p+1}^{N-1} \frac{\partial^2 \ln P(\xi_n | \mathbf{a})}{\partial a_i \partial a_k} \right]_{i,k=1 \ldots p}. \tag{14}$$

To simplify the analysis let's consider the asymptotic estimation of FIM (14) at $N \to \infty$. As the result

$$\mathbf{F} = \mathbf{R} + \mathbf{W} = \mathbf{R} + \begin{bmatrix} w_{0,0} & \mathbf{w}^T \\ \mathbf{w} & \mathbf{W}_{p-1} \end{bmatrix}, \tag{15}$$

where $\mathbf{R} = [r_{|i-k|}]_{i,k=0 \ldots p-1}$ – Toeplitz correlation matrix,

$$w_{0,0} = a_1^2 r_0 + 4 a_1 \sum_{i=1}^{p} a_i r_{i-1} + \sum_{l,m=1}^{p} a_l a_m r_{|l-m|}$$

$$-2\sum_{i=2}^{p} a_i r_{i-2} - 2r_2 ;$$

$$\mathbf{w} = \left[ a_1^2 r_i + 2a_1 \sum_{m=1}^{p} a_m r_{m-1} - a_i r_{i-2} \right]_{i=1...p-1}$$

$$- \left[ \sum_{m=1}^{p} a_m r_{m-i+1} + \sum_{m=2}^{p} a_m r_{m-i-1} \right]_{i=1...p-1} ;$$

$$\mathbf{W}_{p-1} = [r_{|k-i|}(1+a_1^2) - a_1(r_{|k-i-1|} + r_{|k-i+1|})]_{i,k=1...p-1} .$$

From expression (15) it is evident that the system of the equations (13) contributes the additional component matrix $\mathbf{W}$ in the estimation of FIM. If matrix $\mathbf{W}$ is positively determined under some relations between elements of the correlation vector $\mathbf{r}$ and the parameters vector $\mathbf{a}$ then it enlarges energy of FIM and, thereby, reduces KRB. According to this condition the diagonal elements of the matrix $\mathbf{W}$ must be positive. Let us consider main diagonal elements of the matrix $\mathbf{W}$. The element $w_{0,0} > 0$ since $w_{0,0}$ contains dominant component $6a_1^2 r_0$, where $a_1$ is equal to sum of LSO eigenvalues and can not be a small value. Diagonal elements of the matrix $\mathbf{W}_{p-1}$ are positive if function [11]

$$J = r_0(1+a_1^2) - 2r_1 a_1 > 0 . \qquad (16)$$

The function (16) can be used as criterion of the signal dynamical invariance. Positive value of the function points that the system of the equations (13) is consistent.

which are correspond to signal and which are correspond to noise. We will consider the problem of order selection accordingly to the structure of correlation matrix and its relation with singular values distribution (SVD). As it can be seen from expression (12), the given estimation of noise dispersion is minimal and has a physical sense if $\rho_{p-1,p-1}$ is found in neighborhood of positive maximum of the function $\rho_{(p-1)}$. The numerical conditionality of matrix is characterized by its minimal singular value. Let us compare the function $\rho_{(p-1)}$ and SVD using the numeric sequence from work [5] as an example. Given test signal contains sum of three sinusoids and colored noise, volume of the samples set $N = 64$. Authors of the work have chosen the order of the AR as $p = 16$. Figure 1 shows the function $\rho_{(15)}$ and corresponding SVD of the correlation matrix. Numerical experiments have shown that order, which allows to execute inversion of the correlation matrixes is $p = 33$. Function $\rho_{(32)}$ and SVD are shown on figure 2. As it can be seen from figure 2, the second maximum of the function $\rho_{(32)}$ agrees with the biggest dip of the SVD. From figure 2 it is possible to make conclusion that maximal optimum order of the model is $p_{opt\,max} \leq 27$. Maximum of the function $\rho_{(15)}$ does not agreed with dip of the SVD. The examples of spectral estimation in [5] had not sufficient resolution ability on frequency at such order. The function $\rho_{(p-1)}$ has the form on figure 1 at the order $p < 16$ and takes the form similar to submitted on figure 2 at minimal order $p = 24$, then its second maximum and the dip of the SVD are located on the 19th position, consequently $p_{opt\,min} \geq 19$. We will note that the form of the function $\rho_{(p-1)}$ was not greatly changed after addition or

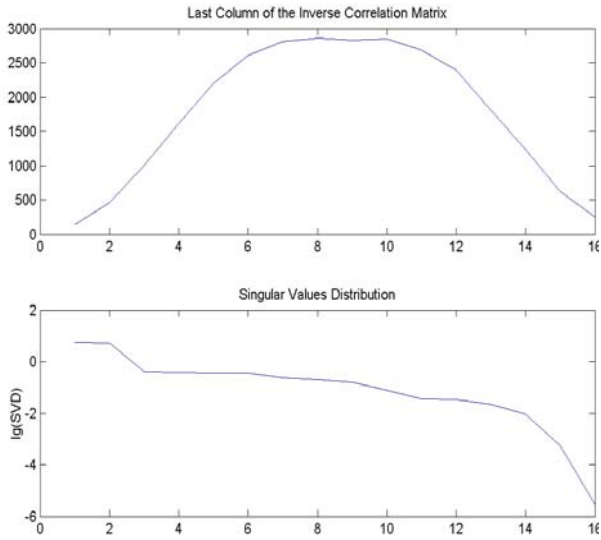

**Figure 1:** Function $\rho_{(15)}$ and SVD at $p = 16$.

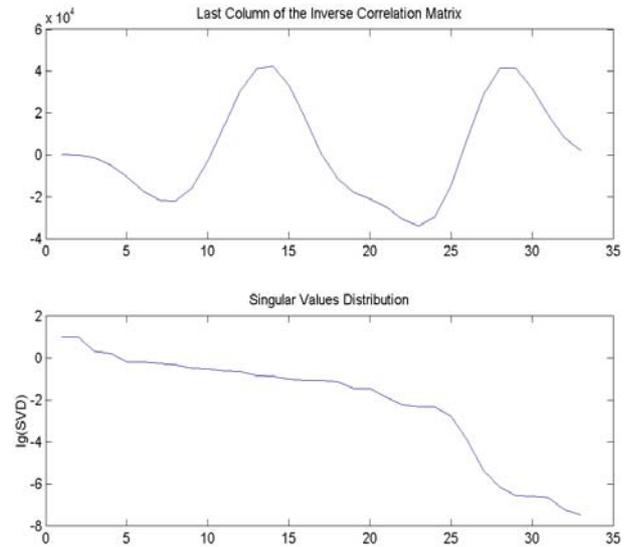

**Figure 2:** Function $\rho_{(32)}$ and SVD at $p = 33$.

### 4. The optimum order of the AR model

Known methods of the AR model order selection as criterion use a minimum value of noise dispersion in equation (1) and are based on Akaike's approach [12]. This criterion suits for order estimation of the second type models and as a rule it gives overfit model [6,13] that is not appropriate for considered problem. Some others approaches are based on the analysis of correlation matrix eigen or singular structures [14-17]. The main idea is based on the separation of the correlation matrix eigen (singular) values on that removing of several sinusoids. A number of greatest singular values points to a number of sinusoids, but in this case the form of the function $\rho_{(p-1)}$ and SVD were defined by colored noise.

The given experiment has shown that for problem of signals location it is necessary to choose the minimum optimum order of the AR model, such that the function $\rho_{(p-1)}$ has two positive maximums, which are separated by value that is negative or about zero. This approach is not universal and is appropriate for determination of the model order of two-component signals as a sum of deterministic signal and noise. For instance, coefficients of the polynomial, which roots are located on the unit circle, can be positive and form the function $\rho_{(p-1)}$ with one maximum.

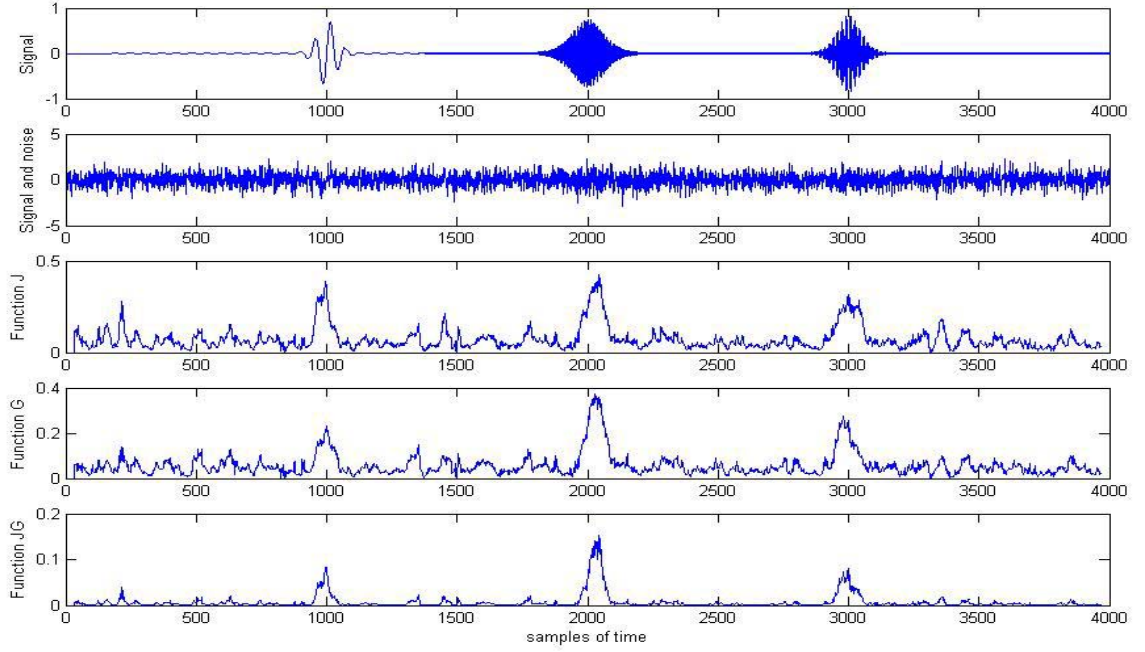

**Figure 3.** Location of wave packets on noise background at SNR $\approx$ -10 dB, $N = 64$.

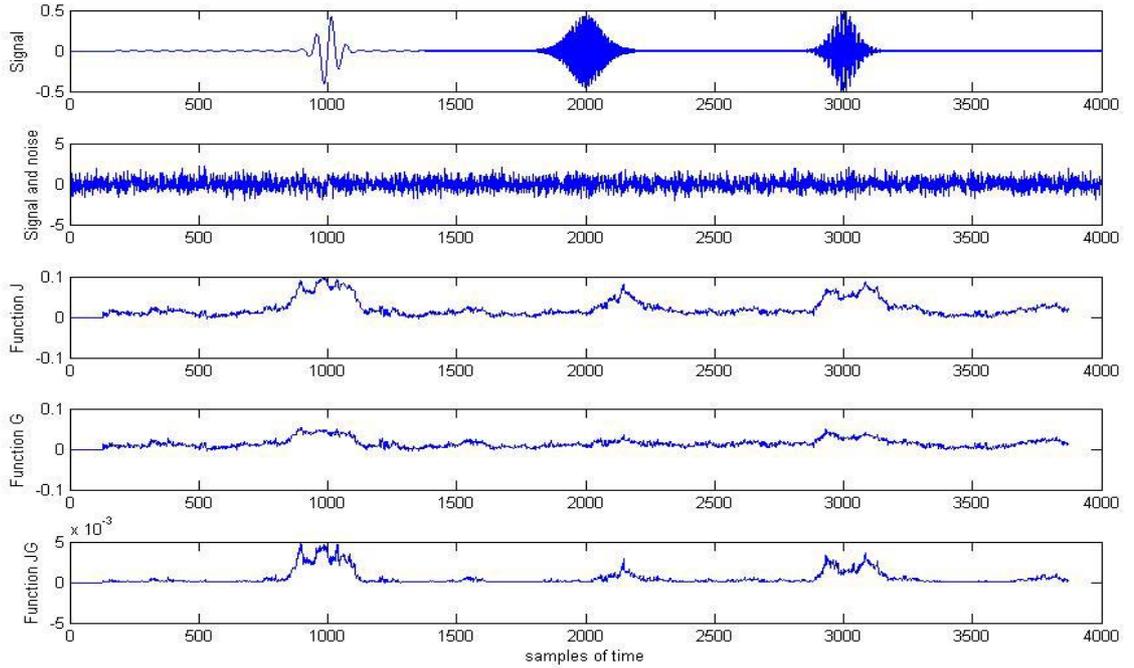

**Figure 4.** Location of wave packets on noise background at SNR $\approx$ -20 dB, $N = 256$.

## 5. Examples of finding wave packets in noise

We have offered two functions to solve the problem of signals finding on noise background. The function $G$ (3) displays power changes of a sought signal. The function $J$ (15) displays presence of dynamical characteristic of invariance with respect to the AR model. Values of the functions must be agreed when signal is present. Therefore we will enter the third function, which will display logical "and" of the first two: $JG = J \cdot G$. Let us write expression for this function with account that power of the signal (2) is possible to be estimated as $P_x = r_{0,0}$ and in expression (15) we shall take into account the influence of noise, in total

$$JG = ((r_0 - \sigma_\xi^2)(1 + a_1^2) - 2r_1 a_1)(r_0 - \sigma_\xi^2), \qquad (17)$$

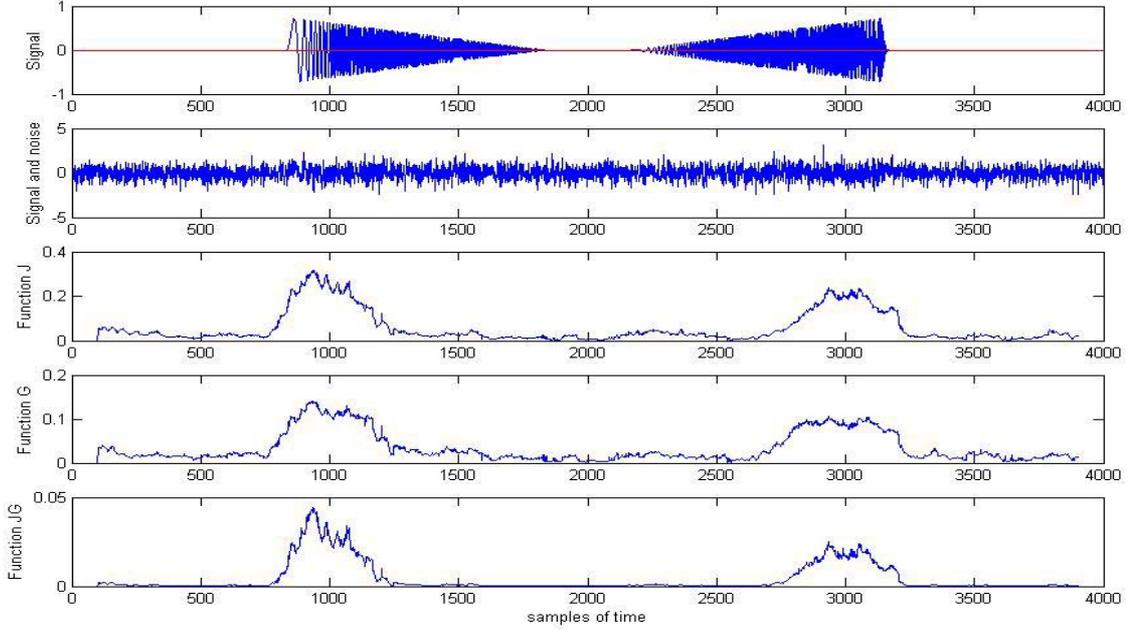

**Figure 5.** Chirp-signals location at SNR $\approx$ -10 dB, $N = 200$.

where $r_0 = r_{0,0}$ and $r_1 = r_{0,1}$, noise dispersion is determined accordingly to the equation (12), the parameter $a_1$ - accordingly to (10).

An example of finding three wave packets is presented on figure 3. The relative frequencies of filling sinusoids are following: 0.018, 0.27, 0.47. The SNR was evaluated on intervals, where signals amplitude was more than 0.1 of maximum, the SNR $\approx$ -10 dB. The noise on statistical characteristic is close to Gauss white. The process of finding was realized as follows. In the beginning the optimum averaged order was calculated for the whole set of data samples, for example on figure 3 $p_{opt} = 8$. Whereupon moving correlation analysis was executed with calculation of the functions $J$, $G$, $JG$ using model with more exact optimal order on each interval. Moving interval of the analysis contains $N = 64$ samples. As it can be seen from figure 3 each packet location is discovered by functions $J$ and $G$, the third function allows to weaken the false surges. The functions $J$ and $G$ doesn't depend from filling frequency. Minimal volume of samples set that allows to locate signal is $N = 48$. Quality of the signals location became much better when $N$ increasing. These characteristics of the method are shown on the second example (figure 4) of finding the same signals at SNR $\approx$ -20 dB and $N = 256$. For evaluation of the function (17) the averaged elements of the correlation matrix were used:

$$r_0 = \frac{1}{p}\sum_{i=0}^{p-1} r_{i,i}\,;\quad r_1 = \frac{1}{p}(\sum_{i=0}^{p-2} r_{i,i+1} + r_{p-1,0})\,.$$

As a result, the values of the functions $J$, $G$, $JG$ of noise component are located in more narrow neighborhood of zero.

The method is appropriate for location of nonstationary signals. Location of two types of chirp-signals is presented on figure 5. This example shows that the signals finding has a probabilistic nature at SNR~-20dB. At the relatively large $N$ there are some mistakes on the borders of signals that can be taken into account.

## 6. Localised high-order spectral analysis

By means of the function $JG$ and the simple condition $JG \geq \varepsilon$, where $\varepsilon$ - certain threshold level, in empirical data it is possible to indicate the areas with useful signal and execute their detailed analysis. Information about location of the signal in noise allows to use known methods of optimum and adaptive filtering that use statistical features apart from the noise and the sum of the signal and the noise [1,2]. We will consider examples of detailed analysis of the presented above test signals using the high order AR (HOAR) model. The HOAR model enables to get approximation of dynamic characteristic of the sum of signal and noise with error dispersion that is far less in comparison with power of the signal. A spectral feature on the basis of the HOAR model allows to define structure of the signal, its discrete and broadband components. The AR model displays frequency characteristic of data and only marginally displays power distribution due to the characteristic polynomial roots of weak components are located closer to the centre of the unit circle. However, the roots of HOAR model can be located in the neighborhood of the unit circle, then spectrum, that is computed by Burg's method of maximum entropy [5,6], has values that are close to singular, a form of spectral peaks of discrete and broadband components is approximately similar. We will use the spectrum estimation on the basis of the forming filter that is invariant to the signal dynamic [18] to avoid distortion of the spectrum. Estimation of the amplitude spectrum has the next form

$$A(f) = \left|\mathbf{L}_p^{-1}(\mathbf{h})\mathbf{e}(f)\right|^{-1}, \qquad (18)$$

where $\mathbf{h}$ - the vector of the pulse transient response, operator of the filter

$$\mathbf{L}_p(\mathbf{h}) = \left[\mathbf{h}^T \quad \mathbf{h}^T \mathbf{K}_p \quad \ldots \quad \mathbf{h}^T \mathbf{K}_p^{p-1}\right]^T, \qquad (19)$$

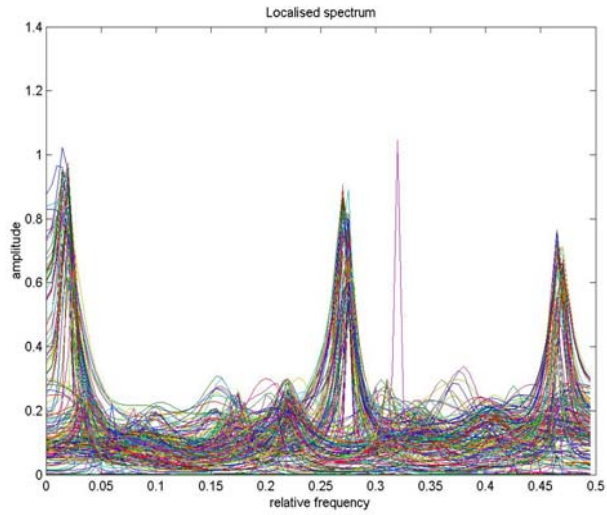

**Figure 6.** Localised spectral estimates at SNR ≈ -10 dB, $N = 64$, $p = 16$.

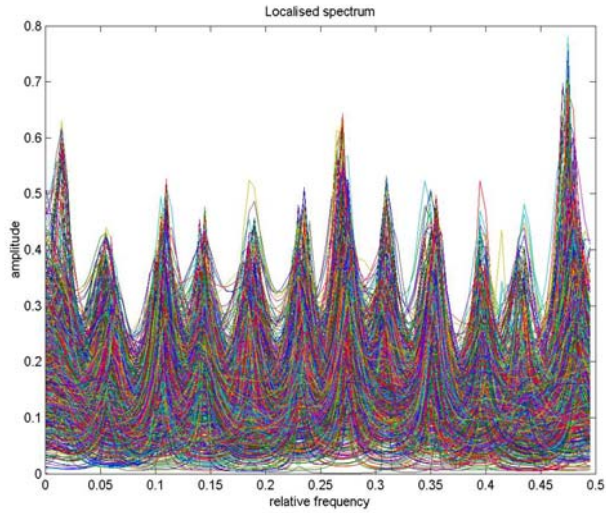

**Figure 7.** Localised spectral estimates at SNR ≈ -20 dB, $N = 256$, $p = 24$.

$$\mathbf{e}(f) = \frac{1}{\sqrt{p}}\left[\exp(2\pi i f k)\right]_{k=0\ldots p-1},$$

relative frequency $f = 0\ldots 0.5$. The vector $\mathbf{h}$ characterizes amplitude spectrum, it can be defined by the the following algorithm. A sum of the LSO eigenvalues is equal to its trace, so samples sequence

$$v_n = tr(\mathbf{K}_p^n) = \sum_{i=1}^{p} z_i^n$$

presents a sum of the LSO eigen harmonic functions. Let us form a matrix $\mathbf{V} = [v_{i+k}]_{i=0\ldots p-1, k=0\ldots N-p}$ and a data vector $\mathbf{x} = [x_i]_{i=0\ldots N-p}$. Each following vector-column $\mathbf{v}_n$ of the matrix $\mathbf{V}$ possible to present as $\mathbf{v}_n = \mathbf{K}_p \mathbf{v}_{n-1}$ so on the basis of the structure of the operator (19) we can write the equation

$$\mathbf{x}^T = \mathbf{h}^T \mathbf{V}.$$

Consequently,

$$\mathbf{h}^T = \mathbf{x}^T \mathbf{V}^{\#},$$

where # - pseudoinversion of a matrix [6].

Estimation of the spectrum using operator method (18) allows to exploit the additional properties of harmonic signals symmetry for reaching more high precision. If during the time interval of analysis the signal is ergodic and stationary then the first type AR model is unitary and doesn't change signal power, therefore the roots of the characteristic polynomial are located on the unit circle. In this case the polynomial is symmetrical in respect of central element and it last coefficient equal to one. Taking into account polynomial symmetry the equation (9) can be presented as following

$$\sum_{i=1}^{q-1} a_i(r_{k,i} + r_{k,p-i} + r_{p-k,i} + r_{p-k,p-i}) +$$
$$a_q(r_{k,q} + r_{p-k,q}) = \quad (20.1)$$
$$\rho_{p-1,p-1}^{-1} \sum_{i=0}^{p-2} \rho_{i,p-1}(r_{k,i+1} + r_{p-k,i+1}) - r_{0,k} - r_{0,p-k}$$

where $p$ is even and $k = 1\ldots q-1$, $q = p/2 - 1$, for $k = q$

$$\sum_{i=1}^{q} a_i(r_{q,i} + r_{i,q} + r_{p-i,q} + r_{q,p-i}) + 2a_q r_{q,q} = .$$
$$(20.2)$$
$$\rho_{p-1,p-1}^{-1} \sum_{i=0}^{p-2}(\rho_{i,p-1} r_{q,i+1} + \rho_{p-1,i} r_{i+1,q}) - r_{q,0} - r_{0,q}$$

The same equation can be found for odd $p$.

Spectral estimations (18) of the first test signal (figure 3) are presented on figure 6. Data fragments are given for analysis by means of the function $JG \geq 0.2 \max(JG)$. The order of the HOAR model $p = 16$, sample size $N = 64$. As it can be seen from figure 6, three maximums are distinctly extracted, its amplitudes and frequencies are located near the real values. Exceeding of maximums over false pulsation is about 30 dB.

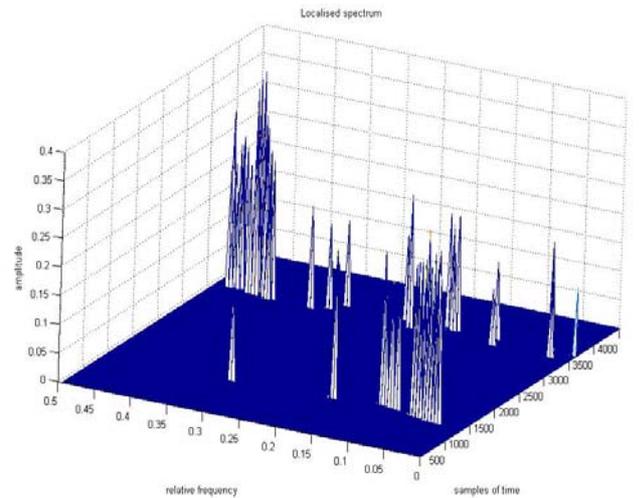

**Figure 8.** Spectral peaks distribution for chirp-signals at SNR ≈ -10 dB, $N = 200$, $p = 16$.

Spectral estimations of the second test signal are presented on figure 7. The order of the HOAR model was increased up to $p = 24$. In this case the power of the noise was distributed on greater number of spectral components. There are three visible maximums which amplitudes and frequencies correspond to real values too. However, in this case excess of the signals power peaks does not exceed 10 dB. But it is not enough to make the certain conclusion about signals parameters and it is necessary to perform some additional processing. So, the considered method can find regular structure of signal when spectral analysis doesn't give satisfactory result.

Frequency and time distribution of spectral peaks of the chirp-signals are presented on figure 8. The order of the HOAR model is $p = 16$ and the sample size is $N = 200$. There are visible erroneous estimates on the signals bounds and peaks lines that reflect signals nature.

## 7. Conclusion

The method of weak short-time signals location in noise is presented. Unlike other methods it is based on the finding of dynamic invariants that are related with AR model of a south signal. Signals dynamic characteristics changes are found with the help of the function (17) that is combination of the functions (3) and (16). The method is devoted to answer the question - does signal contain component with regular dynamic or not?

As the main signal parameter the noise dispersion was used. Noise dispersion was estimated on the same interval that is used for signal finding, without any previous information. This allows to define signals on the nonstationary noise background. Such estimation was reached using characteristic of linear symmetry of the harmonic signal which doesn't have a noise. The characteristic of symmetry is presented by equation of linear symmetry of correlation matrix (6) with respect to action of linear shift operator (5). The given approach in combination with method of the maximum likelihood allows to obtain the estimation of background noise dispersion which is close to real one. Therefore the (12) can serves ratio of data samples power and the estimation of dispersion as a criterion of signal presence or absences. The function (16) has been received using the dynamic invariance characteristic as criterion of model consistency. Joint use of two criteria has allowed to find short wave packets at low value of SNR, when moving analysis of power or spectral analysis can not give satisfactory result.

The method for determination of the optimum order of the AR model was considered in given article. Unlike known methods, recommendations were given for selection of minimum and maximum optimal order. The method is based on the structure of correlation matrix and its relation with SVD, it doesn't depend on the method that was used for the LP parameters estimation.

The high order AR model was used for analysis of found signals nature. The unitary symmetry was used as additional condition for parameters estimation. Spectral estimation (18) with help of the invariant to signal dynamic forming filter (19) was used to avoid singularities of spectral density and to obtain real estimations of the signals amplitudes. The comparison of given estimations with estimations on the basis of the model without characteristic of the unitary symmetry has shown that unitary symmetry allows to get more careful spectrum, which is characterized by clear peaks and smaller scatter of values.

The examples of signals finding have shown that the method doesn't depend from frequency and the required sensitivity can be reached by selection of samples set volume. We shall note that the finding of the fluctuations with one mode presents the most encountered problem, and at the same time the most difficult, since sum of the fluctuations at the same level of noise can be easy found by analysis of the signal power changes.